\documentstyle[preprint,epsfig,aps,12pt]{revtex}
\begin{document}
\draft

\title{Non-linear Poisson-Boltzmann Theory for Swollen Clays}
\author{{\sc R. J. F. Leote de Carvalho (1), 
E. Trizac (1)}\footnote{Present address: 
FOM Institute for Atomic and Molecular Physics,
Kruislaan 407,  1098 SJ Amsterdam, The Netherlands}
and {\sc J.-P. Hansen (2)}}
\address{(1) Laboratoire de Physique, Ecole Normale Sup\'erieure de Lyon 
(URA CNRS 1325) \\
46 All\'ee d'Italie, 69364 Lyon Cedex 07, France \\
(2) Department of Chemistry, University of Cambridge \\
Lensfield Road, Cambridge CB2 1EW, UK}

\date{21 April 1998}

\maketitle

\vspace{1cm}


\begin{abstract}
The non-linear Poisson-Boltzmann (PB) equation for a circular, uniformly 
charged platelet, confined together with co- and counter-ions to a 
cylindrical cell, is solved semi-analytically by transforming it
into an integral equation and solving the latter iteratively.
This method proves efficient and robust, and can be readily 
generalized to other problems based on cell models, treated within 
non-linear Poisson-like theory. The solution to the PB equation
is computed over a wide range of physical conditions, and 
the resulting osmotic equation of state is shown to be
in fair agreement with recent experimental data for 
Laponite clay suspensions, in the concentrated gel phase. 

\end{abstract}


Upon the addition of water, initially dry clays swell into stacks 
of parallel charged platelets separated by increasingly thick layers of
water containing microscopic co- and counter-ions. This simple picture
becomes less realistic as the concentration by weight of clay decreases
and the initial smectic orientational order is gradually lost in
favor of a disordered gel structure. The development of this
gel phase is responsible for characteristic rheological 
properties of the clay dispersion, which are very sensitive 
to salt concentration and of considerable technological importance, 
in particular for the oil drilling industry. Recent experimental work 
\cite{Ramsay,Mourchid,Pignon} focussed on correlating 
the rheological behavior with the
mesoscopic structure of the well-characterized synthetic clay Laponite,
made of quasi-monodisperse disc-like platelets of thickness $\approx
1$nm, diameter $\sigma \approx 30\,$nm, and carrying a negative surface 
charge $-Ze \approx -10^3 \,e$ (where $e$ is the proton charge). The complex
interplay of excluded volume and electrostatic interactions between
such highly anisotropic particles makes a theoretical description of clay 
gels extremely difficult \cite{Dijkstra}. However, the early 
stages of swelling are described reasonably well by a cell model where
each platelet, assumed to be an infinite uniformly charged plane, 
is confined with its monovalent co- and counter-ions
to a slab, and the distribution of the latter is treated within
Poisson-Boltzmann (PB) theory \cite{Dubois}. As swelling proceeds 
the spacing between platelets becomes 
a sizeable fraction of their lateral dimension. Consequently,
edge (finite size) effects become important and the 
problem ceases to be one-dimensional. In the extreme case of 
an isolated platelet immersed in an electrolyte solution (infinite
dilution limit), the PB problem can be solved numerically
\cite{Secor,Chang}. So far, these results provide the only 
available data for disc-like clays and no solution could be
obtained for finite concentrations of platelets.
In this Letter we report results for a cylindrical cell model, 
which accounts for both the finite size of the platelets and 
their finite concentration, and allows a semi-analytical treatment 
within non-linear PB theory. The {\em linearized} version
of PB theory (LPB) for a circular platelet confined with its co- 
and counter-ions to the same cell has been recently considered in 
detail \cite{Trizac}; however, LPB theory is not suited
for large surface potentials, typically for $Z > 10^2$, 
which is an order of magnitude down from physical Laponite charges,
where linearization is no longer justified.

Motivated by the results of \cite{Pignon}, we introduce the 
Wigner-Seitz cell around a clay platelet, and we consider
that the {\em average} shape of this confining cell is
cylindrical in the gel phase. If $n$ is the number 
concentration of platelets, the volume of the 
cylindrical cell is $v = 1/n = 2 \pi R^2 h$ where $R$ is the radius
and $2h$ the height, which is the spacing between platelets in the
stack (typically of the order of 10 nm; however $h$ is not
{\it a priori} prescribed in our model). The platelets 
are assumed to be infinitely thin, of radius
$r_0$, and carrying a uniform surface charge density 
$\sigma = - Ze / \pi r_0^2$. Let $\rho^+({\bf r})$ and $\rho^-({\bf r})$
denote the counter-ion and co-ion density profiles inside the cell. The
electrostatic potential $\varphi({\bf r})$ satisfies Poisson's equation:
\begin{equation}
\label{Poisson}
\nabla^2 \varphi({\bf r}) = - \frac{4 \pi} {\epsilon} \left( q_P({\bf r})
+ e \left[\rho^+({\bf r}) - \rho^-({\bf r})\right]\right)
\end{equation}
where $q_P({\bf r}) = \sigma \delta(z) \theta(r_0 - r)$ is the 
charge density of the platelet in cylindrical
coordinates, and $\epsilon$ is the macroscopic dielectric
constant of water ({\em primitive model}). Within mean-field,
or PB theory, the density profiles are approximated by the Boltzmann
factors:
\begin{equation}
\label{Boltzmann}
\rho^\pm({\bf r})=\rho^\pm_0 \exp \left[ \mp \beta e \varphi({\bf r})\right]
\end{equation}
where $\beta \equiv 1/k_B T$ is the inverse temperature. 
The pre-factors $\rho^\pm_0$ have individually no physical significance,
while the product $\rho^+_0 \rho^-_0$ depends on the conditions under
consideration. If the salt concentration $n_S$ is assumed to be fixed 
in the dispersion, then the concentration of co- and counter-ions in the cell
are imposed: 
$n_-=N_-/{v} = n_S$ and $n_+=N_+/{v} = 
n_S+Z/{v}$, and the
$\rho^\pm_0$ are determined by the normalization constraints (canonical
description):
\begin{equation}
\frac{1}{v} \int_{v} \rho^\pm({\bf r}) d{\bf r} = n_\pm
\end{equation}
On the other hand, if the dispersion is in equilibrium with an ionic solution
of concentration $n'_S$, which acts as a reservoir, then 
the sum of the 
electrochemical potentials of co- and counter-ions 
in the cell must equal the same quantity
for the salt in the reservoir. Since, for consistency, the ions in the cell
are assumed to behave as an ideal solution, chemical equilibrium
implies that \cite{Dubois} (semi-grand-canonical description):
\begin{equation}
\rho_0^+ \rho_0^- = \left(n'_S\right)^2
\end{equation}

By substituting (\ref{Boltzmann}) into (\ref{Poisson}), and subtracting
$\kappa^2 \varphi({\bf r})$ (where $\kappa$ is an arbitrary inverse
length), we arrive at the closed non-linear partial differential
equation for the potential $\varphi({\bf r})$:
\begin{equation}
\label{PB}
\left( \nabla^2 - \kappa^2 \right) \varphi({\bf r}) =  
-\frac{4 \pi}{\epsilon} 
\left( q_P({\bf r}) + e \rho_0^+ {\rm e}^{-\beta e \varphi({\bf r})}
- e \rho_0^- {\rm e}^{\beta e \varphi({\bf r})}
+ \frac{\epsilon \kappa^2}{4 \pi} \varphi({\bf r}) \right)
\end{equation}
Bearing upon the physical significance of the cell model, it 
is assumed that the normal component of the electric field vanishes
everywhere on the surface $\Sigma$ bounding the cylinder cell. Therefore, 
the boundary condition for the potential $\varphi({\bf r})$ is:
\begin{equation}
\label{BC}
\left[ {\bf n}({\bf r}) \!\cdot\! 
\bbox{\nabla}\right]_{{\bf r}\in \Sigma} 
\varphi({\bf r})= 0
\end{equation}
where ${\bf n}({\bf r})$ denotes the normal to the surface
$\Sigma$ at point ${\bf r}$.
The solution of (\ref{PB}) satisfies:
\begin{equation}
\label{varphi}
\varphi({\bf r}) = - \frac{4 \pi}{\epsilon} \int G^\kappa ({\bf r},{\bf r}')
\left( q_P({\bf r'}) +e \rho_0^+ {\rm e}^{-\beta e \varphi({\bf r'})}
- e \rho_0^- {\rm e}^{\beta e \varphi({\bf r'})} 
+ \frac{\epsilon \kappa^2}{4 \pi} \varphi({\bf r}') \right)
d{\bf r}'
\end{equation}
where $G^\kappa ({\bf r},{\bf r}')$ is the Green's function which solves
the linear problem:
\begin{equation}
\label{GF}
\left(\nabla^2 - \kappa^2\right)\, G^\kappa ({\bf r},{\bf r}') = 
\delta({\bf r}-{\bf r}')
\end{equation}
subject to the same boundary condition (\ref{BC}). The reason for replacing
the bare Laplace operator by the screened Laplace operator is that for
$\kappa = 0$ the boundary condition (\ref{BC}) cannot be satisfied
by the solution of eq. (\ref{GF}), as becomes clear by
integrating both sides of eq. (\ref{GF}) (with $\kappa=0$) over the 
cell volume $v$. The Neuman-like boundary condition allows the
Green's function to be expanded, in cylindrical coordinates, 
in the form of a Bessel-Dini series, along the lines of 
the procedure for obtaining the solutions to 
the LPB problems in \cite{Trizac}. We find:
\begin{equation}
\label{green}
G^\kappa ({\bf r},{\bf r}') = 
\sum_{n \geq 1} {\cal{C}}^\pm_n (\phi,{\bf r}')
\cosh \left[ \frac{h\mp z}{\Lambda_n} \right] 
J_0\left(y_n \frac{\rho}{R}\right)
\end{equation}
where ${\bf r} = (\rho, \phi,z)$, the $+$ and $-$ signs correspond to the 
situations $z > z'$ and $z < z'$ respectively, 
$y_n$ is the n$^{th}$ root of $J_1(y)=0$,
$J_0$ and $J_1$ are the zeroth and first order Bessel functions, 
$\Lambda_n^{-2} = (y_n^2 + \kappa^2 R^2) / R^2$, and:
\begin{equation}
{\cal{C}}^\pm_n(\phi,{\bf r}') = 
- \frac{2 \Lambda_n J_0 \left( y_n\, \rho'/R \right) \delta(\phi-\phi')}
{R^2J_0^2(y_n)\sinh \left[2h/\Lambda_n \right]}
\cosh\left[\frac{h\pm z'}{\Lambda_n} \right]
\end{equation}
The kernel $G^\kappa$ being explicitly known, the non-linear integral
equation (\ref{varphi}) may be solved for $\varphi({\bf r})$ by an iterative
Picard procedure starting from an initial guess of $\varphi({\bf r})$
(in most cases we have chosen either $\varphi({\bf r}) \equiv 0$
or the solution of LPB theory \cite{Trizac}). The arbitrary
inverse length $\kappa$ has generally been chosen in the range
$[\kappa_{_{D}}/10,10\,\kappa_{_{D}}]$, $\kappa_{_{D}}$ being the inverse
Debye length in the reservoir. Cylindrical symmetry
implies $\varphi({\bf r})=\varphi(\rho,z)$, which is calculated on a 
two-dimensional $n_\rho \times n_z$ grid spanning half the cylinder, with 
$n_\rho=200$ grid 
points to cover the interval $[0,R]$ (this must be sufficiently
large to allow for a proper representation of the Bessel functions), 
and $n_z=40$ points for the interval $[0,h]$. 50 terms were retained in 
the Bessel-Dini series (\ref{green}), and at each step of the iteration
typically 10\% to 5\% of the {\em new} estimate of $\varphi ({\bf r})$
were mixed with 90\% to 95\% of the previous estimate to ensure proper
convergence; the latter was generally achieved in about 50 to 200 iterations
to ensure a relative accuracy of at least $10^{-5}$. Examples of the resulting
density profiles are shown in Fig. 1 and compared with the LPB predictions;
under {\em realistic} conditions the deviations of LPB from PB theory
are indeed considerable. Provided the iterative procedure
converges, the method yields a solution independent of
$\kappa$, {\em i.e.} no limit $\kappa\to 0$ is
required at the end. We have successfully implemented two independent
checks for the accuracy of the computed potential. 
First, eq. (\ref{varphi}) must be satisfied for any values
of $\kappa$. Secondly, since the multipoles of the total 
charge distribution inside the cell can be transformed
from volume integrals ({\em i.e.} their definition) into surface 
integrals (making use of Poisson's equation), these two 
formulae must yield the same result provided the potential 
$\varphi({\bf r})$ is a solution of the PB problem.

Once the potential $\varphi({\bf r})$ is known for different surface
charge densities $\sigma$, the Helmholtz free energy $F$ can be
calculated by a charging process, {\em e.g.} at constant Debye length
\cite{Trizac}. In the semi-grand-canonical calculations, the number 
of co- and counter-ions, $N_-$ and $N_+$, are computed from
the density profiles (for a given salt concentration $n'_S$
in the reservoir), and the grand potential $\Omega$ is estimated from:
\begin{equation}
\Omega = F - 2 \, N_-  k_B T \ln \left[ n'_S \lambda^3 \right]
\end{equation}
where $\lambda$ is the de Broglie thermal wavelength.

The total quadrupole moment $Q$ of the charge distribution in the cell
(first non-vanishing multipole), the osmotic pressure $\Pi$ and the 
disjoining pressure $\Pi_d$, 
can also be calculated from the density profiles as
described in ref. \cite{Trizac}. For a given
clay concentration $n$, and hence cell volume $v$, all these
quantities depend on the aspect ratio $h/R$ (or equivalently
$h/r_0$) of the cell. The equilibrium stacking is determined by 
minimizing either the free energy or the grand potential with respect 
to $h/r_0$. This is illustrated in Fig. 2. 
As expected, $h$ is found to be of the order of 10 nm. The minima 
from the PB calculations turn out to be consistently flatter than the
LPB results, pointing to the possibility of large fluctuations
of the stacking configurations. The optimum aspect ratios
$h/r_0$ predicted by PB and LPB are practically identical. This ratio 
varies with $n$, but is very insensitive to salt concentration.
As in the LPB case, $Q$ is found to vanish at the minimum
of the relevant thermodynamic potential
($F$ in the canonical case or $\Omega$ in a semi-grand-canonical
description, the latter being suited for a comparison with the
experiments of \cite{Mourchid}), while $\Pi$ and $\Pi_d$ are equal 
at that point. This property is exact and may serve as a 
check for the consistency of the numerical calculations,
while the vanishing of $Q$ cannot be proven {\it a priori} within
the PB framework, but may be regarded as an indication of the validity of
our cell model for the description of dispersions of
clay platelets \cite{Trizac}. 

Our PB results for the osmotic equation of state, $\Pi$ versus
clay concentration, are compared in Fig. 3 to the experimental data
of Mourchid et al. \cite{Mourchid} for three reservoir salt concentrations
$n'_S$. The agreement is reasonable at the higher concentrations
(above $C \approx$ 4\% $(w/w)$ in units of \% of weight
of Laponite per weight of solvent (water)) corresponding 
to the gel phase, except for the lowest 
salinity investigated ($n_S'=10^{-4}$M with 1M$ = 1 $mol dm$^{-3}$)
for which the convergence criterion is more difficult to meet.
At this low salinity the PB predictions overestimate the osmotic
pressure considerably, compared to the experiments. This disagreement
may be linked to an excessive depletion of salt (Donnan effect)
predicted by PB theory (see Table 1). Too strong a depletion
leads to insufficient screening, and hence to an overestimation
of the osmotic pressure.

The cell model should not apply at low clay concentrations where 
the parallel stacking of platelets is completely lost,
giving rise to the orientationally disordered sol phase.
Since there is no long-ranged orientational order in the gel phase 
\cite{Mourchid,Pignon} our cell model tends to overestimate
the effect of lamellar order. 
However, the analytical results 
obtained in \cite{Trizac} within LPB indicate that given a clay concentration,
and at least at this level of approximation, the precise shape of 
the confining cell does not affect significantly mesoscopic 
or macroscopic quantities such as the osmotic pressure.
Finally, note that LPB theory 
leads to negative osmotic pressures under most conditions 
represented in the figure. 

In order to assess the importance of edge effects we
have also considered
the case of infinite planar platelets having the same charge and mass
density than the finite Laponite discs. The corresponding one-dimensional
PB problem can be solved numerically along the lines in 
ref. \cite{Dubois}. For a given concentration by weight in Laponite
the spacing $2h_\infty$ between the infinite
platelets differs from the optimum (lowest free energy) spacing
between the finite discs (which is practically independent of
$n'_S$). A comparison is made in Fig. 3. It is clear that
as $n'_S$ increases edge effects become more important and the 
simple model of a stack of charged layers cannot describe
the structure of the Laponite suspension. The two curves
for the osmotic pressure as obtained from this simple model 
at the higher concentrations $n'_S$ are completely off scale.
In view of the poor performance of PB theory for finite platelets
at $n'_S = 10^{-4}$M, the reasonable agreement of
the osmotic pressure calculated for infinite platelets with experimental
data must be considered as fortuitous.

In summary, the present PB calculations confirm the qualitative
features of earlier LPB results but while the
latter predictions for the osmotic pressure are unphysical, the PB 
results are in reasonable agreement with experimental data for Laponite
in the high concentration regime. The transformation of the initial
Poisson equation into an integral equation problem, combined with 
the analytical calculation of the corresponding Green's function,
proves efficient and can be readily generalized to other Poisson 
problems with non-linear source terms.

We thank Thierry Biben for many stimulating
discussions. RJFLdC is presently 
carrying out work at the ENS de Lyon as part of a project financed 
by the European Commission through the Training and Mobility of 
Researchers (TMR) programme. ET acknowledges the hospitality 
of Professor Daan Frenkel (AMOLF, Amsterdam).

\begin{table}
\vspace*{4cm}
\begin{minipage}{\textwidth}
\begin{center}
\begin{tabular}{c|cccc}
&
\multicolumn{4}{c}{Values of $n_S / n'_S$ at} \\
$C$ \% $(w/w)$ &
$n'_S = 10^{-4}$M &
$n'_S = 10^{-3}$M &
$n'_S = 5\times 10^{-3}$M & 
$n'_S = 10^{-2}$M \\
\hline
 3.293 & 0.025 & 0.238  & 0.654 & 0.786 \\
 4.390 & 0.018 & 0.168  & 0.560 & 0.716 \\
 5.488 & 0.013 & 0.125  & 0.476 & 0.653 \\
\end{tabular}
\end{center}
\end{minipage}
\vspace{0.1cm}
\caption{Values for the ratio between the salt concentration
in the system, $n_S$, and in the reservoir, $n'_S$, for
several Laponite concentrations. These are 
obtained from semi-grand canonical non-linear PB
calculations, at the aspect ratio that 
minimizes the grand potential.}
\end{table}

\begin{figure}
\vspace*{1cm}
\centerline{\psfig{figure=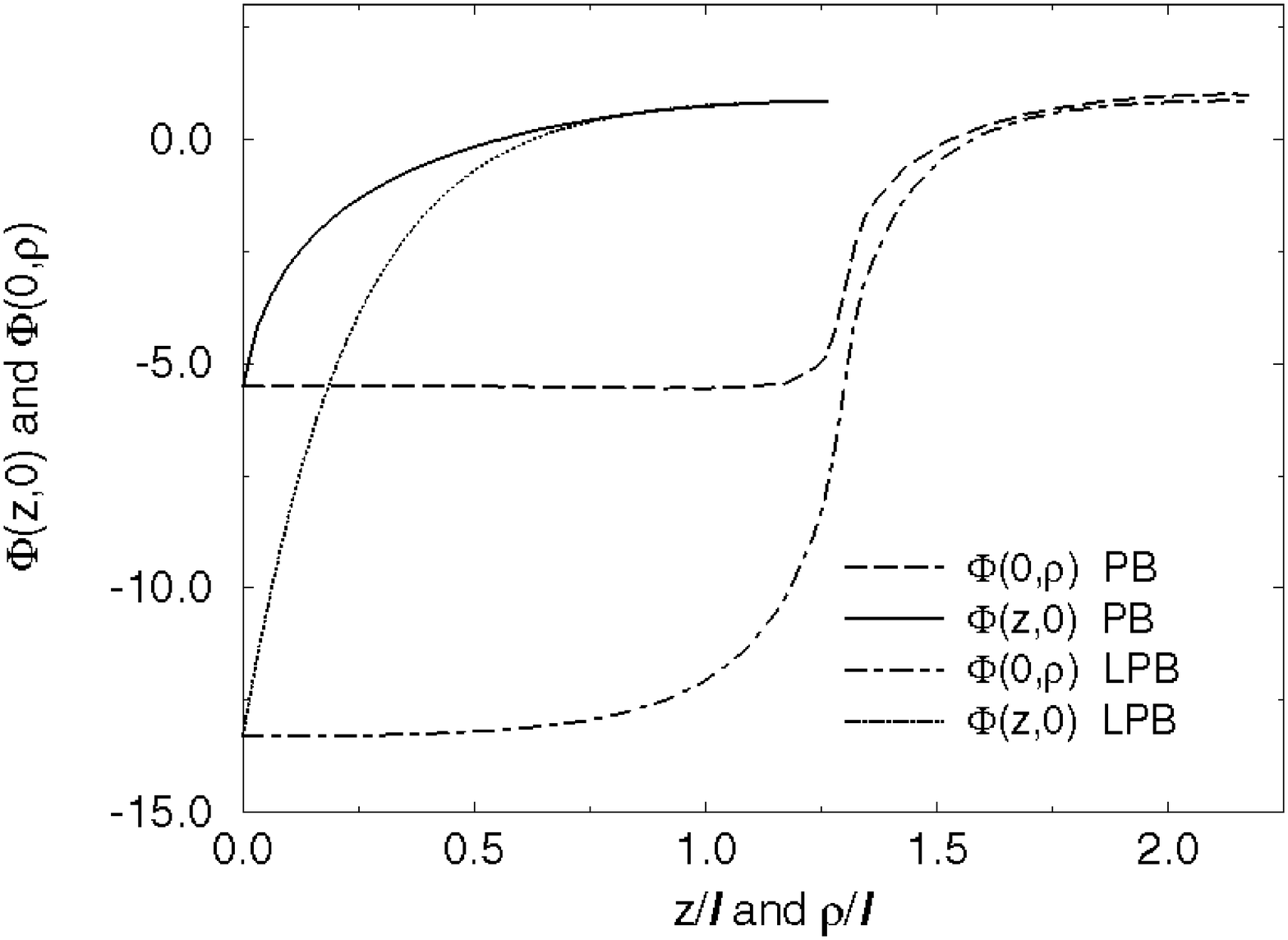,width=\textwidth}}
\caption{Dimensionless electrostatic potential 
$\Phi(z,\rho) \equiv \beta e \varphi(z,\rho)$ 
at $z=0$ versus $\rho/\ell$, 
and at $\rho=0$ versus $z/\ell$,
with $\ell^{-2} = 4 \pi \beta e^2 (2 n_S)/ \epsilon_0\epsilon_r$ . 
The profiles were obtained in non-linear PB, with
the calculations performed in the canonical ensemble, and
in LPB (see \protect\cite{Trizac} for details).
The results in this figure are for an aqueous 
solution ($\epsilon_r = 78$)
of clay discs of radius $r_0 = 125$ \AA$\,$and
surface charge $Z = 700$, at a temperature $T = 300$ K.
The concentration of clay is $n = 5 \times 10^{-5}$M 
and that of added monovalent salt $n_S = 10^{-3}$M.
The aspect ratio is $h/r_0=0.971$, the value that minimizes
the Helmholtz free energy (see Fig 2).}
\end{figure}
\begin{figure}
\centerline{\hspace*{-2cm}\psfig{figure=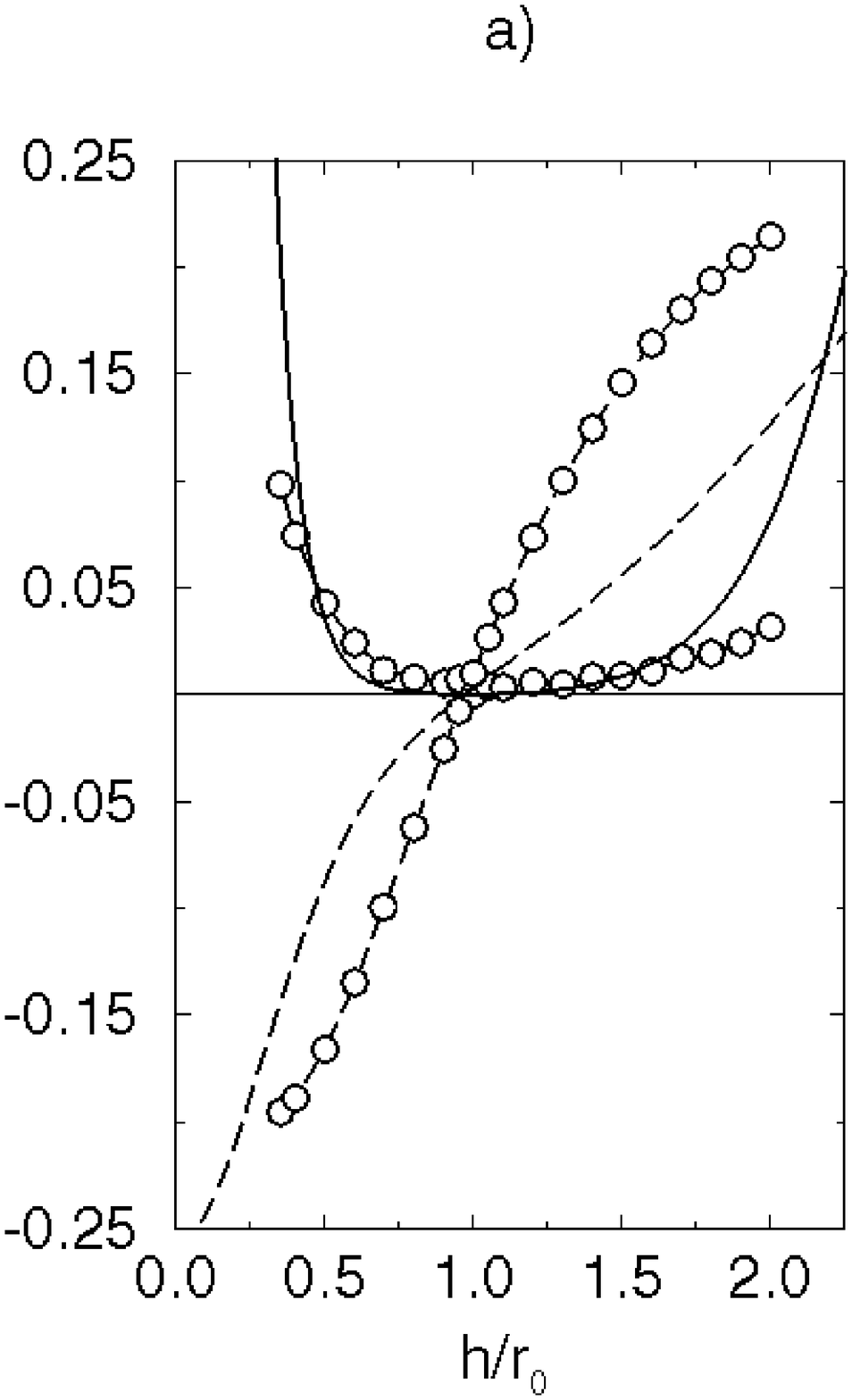,width=\textwidth}}
\vspace*{-12.75cm}
\centerline{\hspace*{13cm}\psfig{figure=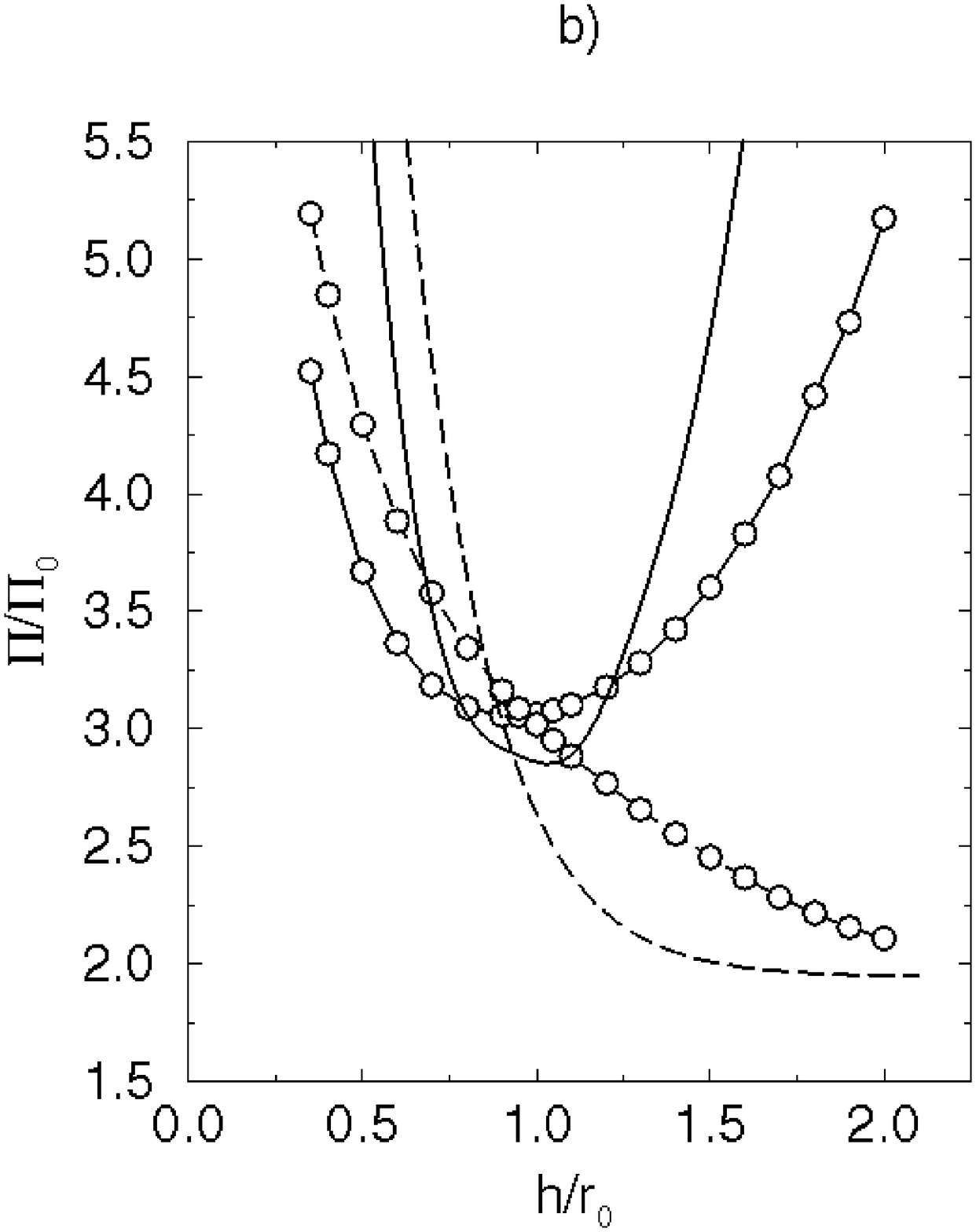,width=\textwidth}}
\caption{a) Dependence of the dimensionless
Helmholtz free energy $\beta (F - F_0) / Z$
and the dimensionless quadrupole moment 
$Q/Q^{disc}$, with $Q^{disc} = Ze r_0^2 / 4$ and
$Q = Q_{zz} =$ $ -2 Q_{xx} =$ $ -2 Q_{yy}$, on the aspect ratio
$h/r_0$ of the Wigner-Seitz cylindrical cell. The solid lines
represent $\beta (F - F_0) / Z$ and the dashed lines are for
$Q/Q^{disc}$. b) Dependence of the
disjoining $\Pi_d$ and osmotic $\Pi$ pressure on $h/r_0$.
The dashed lines correspond to $\Pi_d$ and the solid
lines to $\Pi$, with $\Pi_0 = k_B T (2n_S)$. 
In both plots the curves with $\bigcirc$ were 
obtained from PB calculations whereas
those with no symbols represent LPB results.
At the optimum $h/r_0$, which minimizes $F$, 
$Q$ vanishes and $\Pi = \Pi_d$ holds.
The optimum value is consistently found to be $h/r_0 = 0.971$,
and $F_0$ is the value of $F$ at this aspect ratio.
The canonical ensemble calculations correspond to the same conditions
as in Fig 1. The grand potential calculated in the semi-grand-canonical
ensemble at a reservoir salt concentration
$n_S' = 2.54257 \times 10^{-3}$M exhibits a minimum at the 
same value of $h/r_0$, where $Q$ vanishes and
$\Pi = \Pi_d = 2530$ Pa from the calculations in both ensembles.}
\end{figure}
\begin{figure}
\centerline{\psfig{figure=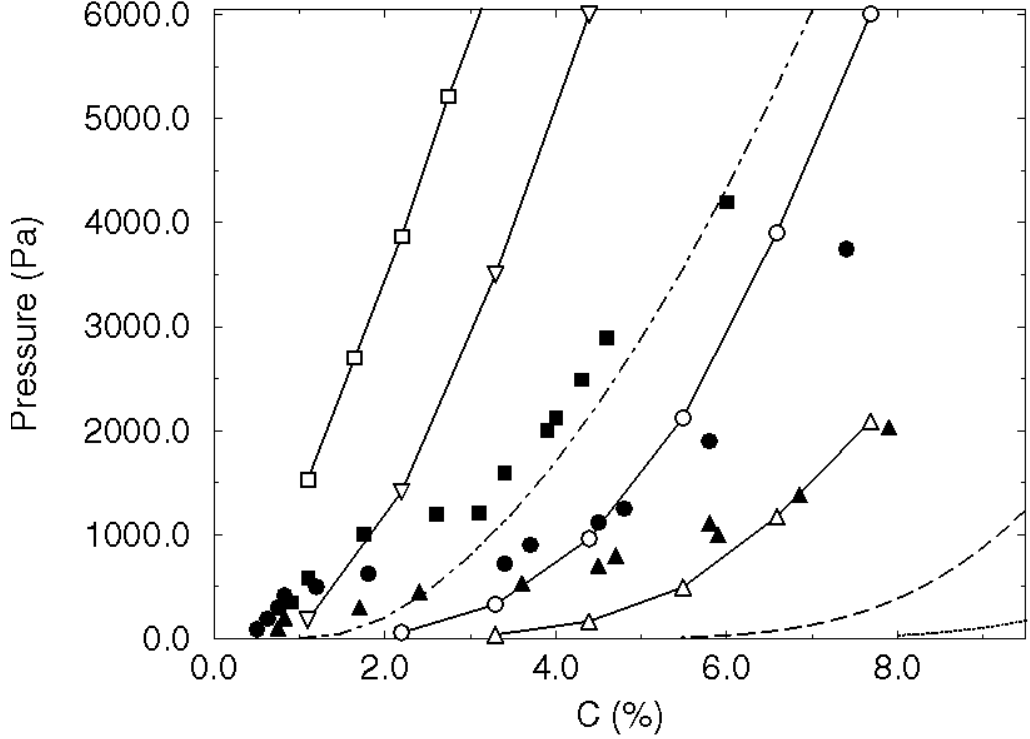,width=\textwidth}}
\caption{Equations of state of Laponite suspensions for different
reservoir salt concentration $n_S'$. The filled symbols
represent experimental results obtained by Mourchid {\it et al.}
\protect\cite{Mourchid}. The solid lines with open symbols represent
our semi-grand canonical PB results obtained at the same values 
of $n_S'$. Each of these points corresponds to the osmotic
pressure at the minimum of the grand potential
with respect to $h/r_0$, at fixed $n_S'$.
The values of $n_S'$ are
$10^{-4}$M ($\Box$),
$10^{-3}$M ($\bigtriangledown$),
$5 \times 10^{-3}$M ($\bigcirc$) , and
$10^{-2}$M ($\triangle$). The
dash-dotted, the dashed and dotted lines are  
the equations of state obtained from the one-dimensional PB calculation 
for a stack of infinite planar charged layers in the presence
of salt,
with the same reservoir concentration $n_S'$, respectively. 
The computed results are for an
aqueous solution ($\epsilon_r = 78$) of Laponite
at $T = 300$ K. The
radius of the clay discs is $r_0 = 150$ \AA$\,$
and the surface charge $Z =  1000$.}
\end{figure}

\end{document}